\documentclass{iopart}
\usepackage{bm}
\usepackage[dvipdf]{graphicx}
\usepackage{yfonts}
\usepackage{enumerate}
\usepackage{color}
\usepackage{caption}

\newcommand{\bra}{\left\langle}
\newcommand{\ket}{\right\rangle}

\newcommand{\pder}[2]{\frac{\partial #1}{\partial  #2}}

\newcommand{\der}[2]{\frac{d #1}{d  #2}}
\newcommand{\dert}[2]{\frac{d^2 #1}{d  #2^2}}

\newcommand{\pex}{p_{\rm ex}}
\newcommand{\hatpex}{\hat{p}_{\rm ex}}
\newcommand{\D}{\mathcal{D}}

\newcommand{\Ob}{\mathcal{O}}

\newcommand{\IMSR}{I_{\rm MSR}}

\begin{document}

\title{Thermodynamic entropy as a Noether invariant in a Langevin equation}
\author{Yuki Minami$^1$ and Shin-ichi Sasa$^2$}
\address{Department of Physics, Zhejiang University, Hangzhou 310027,
  China$^1$, 
Department of Physics, Kyoto University, Kyoto 606-8502, Japan$^2$}
\eads{yminami@zju.edu.cn$^1$, sasa@scphys.kyoto-u.ac.jp$^2$}
\date{\today}
\pacs{ 
05.70.Ln,
05.40.-a, 
}

\begin{abstract}
  We study the thermodynamic entropy as a Noether invariant in a
stochastic process. Following the Onsager theory, we consider the Langevin
equation for a thermodynamic variable in a thermally isolated system. 
By analyzing the Martin--Siggia--Rose--Janssen--de Dominicis action of
the Langevin equation, we find that this action possesses a
continuous symmetry in quasi-static processes, which leads to the
thermodynamic entropy as the Noether invariant for the symmetry.
\end{abstract}


\section{Introduction}

Symmetry is one of the most important notions in modern physics.
The symmetry reveals non-trivial laws independently of the details
of systems under consideration. Well-known examples are
Noether's theorem~\cite{Noether1918}, the Nambu-Goldstone theorem~\cite{nambu1961dynamical, goldstone1961field, goldstone1962broken}, and the Ward-Takahashi identity~\cite{PhysRev.78.182, takahashi1957generalized}.
Noether's theorem ensures the existence of a conservation law
from a continuous symmetry. The Nambu-Goldstone theorem shows
the appearance of gapless modes associated with a spontaneous
breaking of a continuous symmetry ~\cite{nambu1961dynamical, goldstone1961field, goldstone1962broken, PhysRevLett.110.091601, PhysRevLett.108.251602, PhysRevD.91.056006, minami2018spontaneous}.
The Ward-Takahashi identity provides a relation among correlation functions from a symmetry. 
 
If one assumes the converse of Noether's theorem, the existence of
a conserved quantity may suggest the existence of a continuous symmetry.
From this viewpoint, the thermodynamic entropy, which is a conserved
quantity in quasi-static adiabatic processes, can be characterized as the
Noether invariant associated with a continuous symmetry for  classical
Hamilton systems and quantum systems~\cite{sasa2016thermodynamic,
sasa2019thermodynamical}. According to~\cite{sasa2016thermodynamic}, in isolated classical many-body systems with an external control parameter,
the action possesses a non-uniform time-translation symmetry
\begin{eqnarray}
t \to t + \eta \hbar \beta \label{eq:transSY}
\end{eqnarray}
for a class of phase-space trajectories that correspond
to thermodynamic processes, 
where $\eta$ is an infinitesimal parameter, $\hbar$ is the Planck constant,
and $\beta$ is the inverse temperature that depends on a phase space
point through the statistical mechanical formula.  For isolated quantum many
body systems with an external step protocol parameter, the effective
action in a thermodynamic state space was derived under some physically
reasonable assumptions ~\cite{sasa2019thermodynamical}. This action
possesses a symmetry for a uniform translation of the canonical
conjugate variable of the entropy.   

\begin{figure}[htbp]
  \begin{center}
    \begin{tabular}{c}
      \begin{minipage}{0.5\hsize}
        \begin{center}
          \includegraphics[width=\hsize]{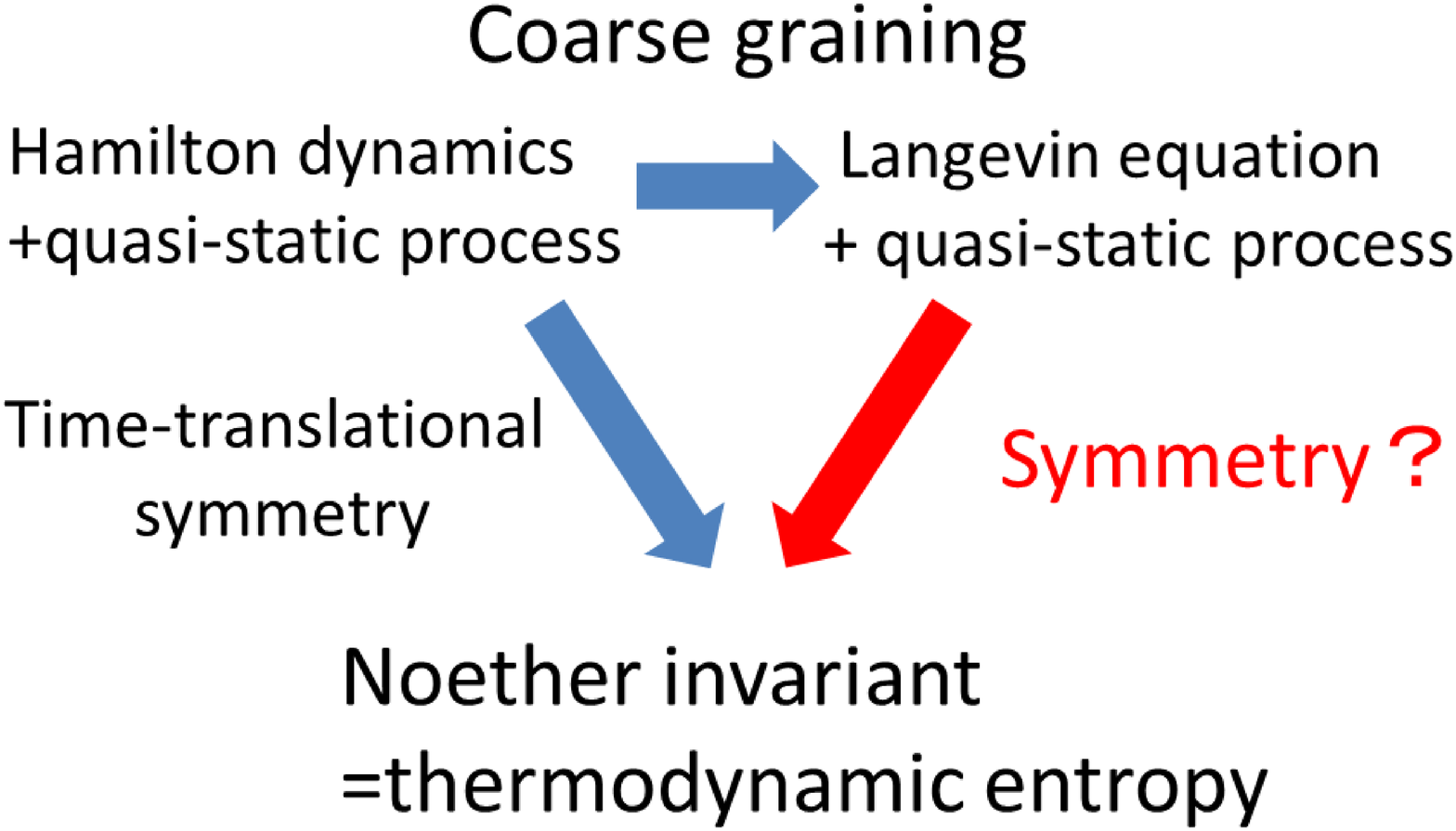}
          \caption{Illustration of the relation among symmetry,
            thermodynamic entropy and coarse graining.\label{fig:motivation} }
        \end{center}
      \end{minipage}
      \begin{minipage}{0.5\hsize}
        \begin{center}
          \includegraphics[width=0.9\hsize]{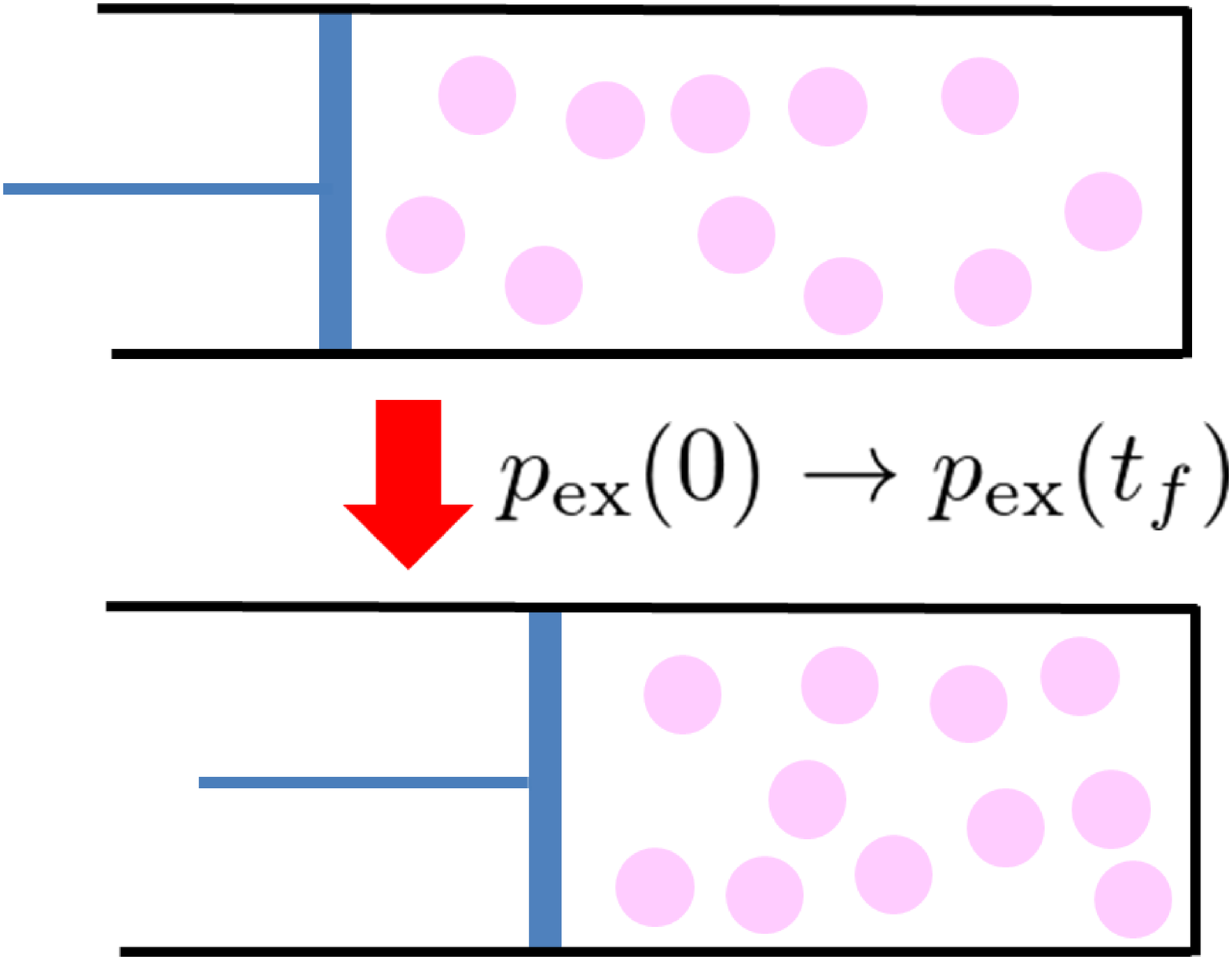}
          \caption{Particles in a container enclosed by thermally isolated
walls. One side is a movable wall under pressure $\pex(t)$. We externally control the pressure as a function of time: $\pex(0) \to \pex(t_f)$.
 \label{fig:setup} }
        \end{center}
      \end{minipage}
    \end{tabular}
  \end{center}
\end{figure}

Here, because Langevin equations may be derived
by a coarse graining method for the microscopic dynamics
~\cite{Zwanzig1961}, a natural question arises:
Can the entropy of a stochastic process be obtained as a
Noether invariant. In this study,  we find a symmetry
by which the entropy is characterized as a Noether
invariant of the Langevin equation that describes a stochastic
process of a thermodynamic variable in thermally isolated systems.
Fig.~\ref{fig:motivation} illustrates our motivation. 

\captionsetup[figure]{font=small,justification=raggedright}

Recall that the fluctuation-dissipation theorems are understood
from the time reversal 
symmetry~\cite{PhysRev.37.405, PhysRev.38.2265, RevModPhys.17.343, kubo1966fluctuation, PhysRevD.30.1218, gao2018emergent}. Similarly, the entropy
conservation in quasi-static processes, which is known as 
the adiabatic theorem, was derived from the Hamiltonian mechanics with
the reversal symmetry \cite{Anosov1960,Kasuga1961,Ott1979,Jarzynski1993}.
In this study, we prove the adiabatic theorem in a stochastic system using a fluctuation theorem similar to the Jarzynski equality~\cite{PhysRevLett.78.2690}, closely related to the time
reversal symmetry. 
Nevertheless,  the entropy
cannot be described as a Noether invariant from the time reversal symmetry because it is discrete.
Therefore, in this study, we explore a continuous symmetry that leads
to such conservation as a result of Noether's theorem. 

As a simple example, we specifically study
particles in a container enclosed by thermally isolated
walls. One wall is movable and externally controlled
by a time-dependent pressure $\pex(t)$, as shown
in Fig.~\ref{fig:setup}. According to thermodynamics,
the total entropy in this system is conserved when the pressure
$\pex(t)$ changes in the quasi-static manner. For this
system, we can describe stochastic processes of the volume
by a Langevin equation. Even in this stochastic description,
the total entropy is conserved in the quasi-static processes. 
We thus formulate a symmetry property of this stochastic system.


Technically, we derive the symmetry by employing
a path integral representation, known as
Martin--Siggia--Rose-Janssen--de Dominicis (MSRJD) formalism~\cite{MSR,J,D1,D2},
to the Langevin equation. The MSRJD
formalism has the similar structure as the path integral formulation
of quantum dynamics, and the theoretical arguments for the symmetry in quantum
dynamics have been developed~\cite{WeinbergText}. Thus, we can use 
techniques in the MSRJD formalism to discuss the symmetry property,
as was done in \cite{minami2018spontaneous,aron2016dynamical,aron2018non}.


The remainder of this paper is organized as follows. 
In Sec.~\ref{setup}, we explain our setup
and provide the stochastic description of the system in question.
Then, we derive a fluctuation theorem for our model. Using the
obtained equality, we show that the entropy is conserved in the
quasi-static processes. Additionally, we provide the MSRJD formalism
of the Langevin equation. 
In Sec.~\ref{sec:symmetry}, we first map the MSRJD formalism to Hamilton mechanics.  Through the study of a Poisson bracket relation, we identify the entropy as a generator that yields a transformation. Then, we investigate whether an action in the path integral representation is symmetric under such transformation or not.
In Sec.~\ref{sec:identity}, we derive the Ward-Takahashi identity
from the transformation generated by the entropy, and find that the
symmetry breaking terms lead to a fluctuation-response relation at equal time. 
We also derive a thermodynamic uncertainty relation as a result of the identity.  
Sec.~\ref{remarks} concludes the paper.

Note that, throughout this paper, we set the Boltzmann constant to be unity.

\section{Setup}\label{setup}

\subsection{Model}

We consider $N$ particles in a container enclosed by thermally isolated
walls. One side is a movable wall under pressure $\pex$,
as shown in Fig.  \ref{fig:setup}.
We first consider the case where $\pex$ is constant. Let $E$ and $V$
be the internal energy of the particles and the volume of the container, while
$S(E,V)$ denotes the entropy function for this system. In this case,
$E$ and $V$ fluctuate while satisfying the energy conservation
\begin{equation}
  E+\pex V =E_{\rm tot}={\rm const}, 
\label{E-con}
\end{equation}
where $E_{\rm tot} $ corresponds to the enthalpy.

According to equilibrium thermodynamics, the probability density of
$V$ is given by
\begin{equation}
P_{\rm eq}(V;E_{\rm tot},\pex)=
                  \exp\left[S(E_{\rm tot}-\pex V, V)-S_*(E_{\rm tot},\pex) \right],
\label{p-eq}
\end{equation}
where $S_*(E_{\rm tot},\pex)$ is determined from the normalization
of the probability. This is estimated as 
\begin{equation}
S_*(E_{\rm tot}, \pex)=\max_{V} S(E_{\rm tot}-\pex V, V)+o(N)
\end{equation}
through the saddle point calculation.
$S_*$ is the thermodynamic entropy as a function of the enthalpy $E_{\rm tot}$
and the pressure $\pex$. Following the standard thermodynamics notation,
$S_*(E_{\rm tot}, \pex)$ is also written as $S(E_{\rm tot}, \pex)$; however we do not
use $S(E_{\rm tot}, \pex)$ to distinguish it from $S(E,V)$.
From the Onsager theory, the time evolution of fluctuating $V$ is written as
\begin{equation}
  \der{V}{t}= L \der{S(E_{\rm tot}-\pex V, V)}{V}+\sqrt{2L} \xi,
\label{ons}
\end{equation}
where $L$ is the Onsager coefficient and $\xi$ is the Gaussian-white noise
satisfying 
\begin{equation}
  \bra \xi(t) \xi(t') \ket =\delta(t-t') .
\end{equation}
The stationary probability density of $V$ is equal to (\ref{p-eq}).

Now, we consider the case where $\pex$ depends on time $t$.
The system is then described by (\ref{ons}) and
\begin{equation}
\der{E_{\rm tot}}{t}= V \der{\pex  }{t},
\label{dotE}
\end{equation}
where the initial value $E_{\rm tot}(0)$ is assumed. 

\subsection{Fluctuation theorem}

For the considered model, we derive a fluctuation theorem  and confirm
the second law of thermodynamics. 
Let $\hat V=(V(t))_{t=0}^{t={t_f}}$
and $\hatpex=(\pex(t))_{t=0}^{t={t_f}}$. Then, for the fixed $V(0)$, $E_{\rm tot}(0)$, and $\hatpex$, the
probability density of trajectories is given by
\begin{eqnarray}
\fl   {\cal P}(\hat V|V(0),E_{\rm tot}(0), \hatpex  )
&=& {\cal N} \exp\left\{
    -\int_0^{t_f} dt \left[
      \frac{1}{4L}
      \left( \der{V}{t}- L \der{S(E_{\rm tot}-\pex V, V)}{V} \right)^2
                     \nonumber \right. \right.  \nonumber \\
&&  \left. \left.
      +
      \frac{1}{2}\dert{S(E_{\rm tot}-\pex V, V)}{V} 
      \right] \right\} ,
\label{OM}
\end{eqnarray}
where ${\cal N}$ is the normalization constant. 
Next, for given $\hat V$, $\hatpex$ and $E(0)$, we define the time-reversed
trajectory $V^\dagger(t)\equiv V({t_f}-t)$, the time reversed protocol
$\pex^\dagger(t)\equiv \pex({t_f}-t)$, and 
\begin{equation}
E^\dagger(0)\equiv  E(0)+ \int_0^{t_f}  dt V(t)\der{\pex(t)}{t}.
\end{equation}
Then, we have 
\begin{eqnarray}
&&  \frac{{\cal P}(\hat V^\dagger |V^\dagger(0),E_{\rm tot}^\dagger(0),
                    \hatpex^\dagger  )}
        { {\cal P}(\hat V|V(0),E_{\rm tot}(0), \hatpex  )} \nonumber \\
&&
 = \exp\left(
    -\int_0^{t_f} dt  \der{V}{t} \der{S(E_{\rm tot}-\pex V, V)}{V}  \right).
\label{10}
\end{eqnarray}
Here, 
\begin{eqnarray}
&&   \der{S(E_{\rm tot}-\pex V, V)}{t}  \nonumber \\
&=&
\der{V}{t} \der{S(E_{\rm tot}-\pex V, V)}{V}
+\pder{S(E_{\rm tot}-\pex V, V)}{E_{\rm tot}}
\left[\der{E_{\rm tot}}{t}-\der{\pex }{t} V \right]  \nonumber \\
&= & 
\der{V}{t} \der{S(E_{\rm tot}-\pex V, V)}{V}
\label{11}
\end{eqnarray}
holds according to (\ref{dotE}). By substituting (\ref{11}) into
(\ref{10}), we obtain
\begin{eqnarray}
   \hspace{-0.5cm}  \frac{{\cal P}(\hat V^\dagger |V^\dagger(0),E_{\rm tot}^\dagger(0),
                    \hatpex^\dagger  )}
         {{\cal P}(\hat V|V(0),E_{\rm tot}(0), \hatpex  )} \nonumber \\
  \hspace{-0.5cm}      =
 \exp\left[
   S(E_{\rm tot}-V(0)\pex(0), V)
   - S(E_{\rm tot}^\dagger(0)-V^\dagger(0)\pex^\dagger(0), V^\dagger(0))
     \right].
\end{eqnarray}
This can be rewritten using (\ref{p-eq}) as
\begin{eqnarray}
&&  \frac{{\cal P}(\hat V^\dagger |V^\dagger(0),E_{\rm tot}^\dagger(0),
    \hatpex^\dagger  )
    P_{\rm eq}(V^\dagger(0);E_{\rm tot}^\dagger(0),\pex^\dagger(0))
    }
  { {\cal P}(\hat V|V(0),E_{\rm tot}(0), \hatpex  )
    P_{\rm eq}(V(0);E_{\rm tot}(0),\pex(0))}
   \nonumber \\
&&
 =
 \exp
 \left[- \left( S_*(E_{\rm tot}(0),\pex(0)) -    S_*(E_{\rm tot}(0),\pex(0))
    \right)
   \right],
\end{eqnarray}
Because the expectation of $ A(\hat V) $ for any function of $\hat V$
is written as
\begin{equation}
  \hspace{-0.5cm}  \bra A \ket \equiv 
\int{\cal D}\hat V   { {\cal P}(\hat V|V(0),E_{\rm tot}(0), \hatpex  )
    P_{\rm eq}(V(0);E_{\rm tot}(0),\pex(0))} A(\hat V),
\end{equation}
we obtain
\begin{equation}
\bra 
 \exp
 \left[- \left( S_*(E_{\rm tot}({t_f}),\pex({t_f}))
   -    S_*(E_{\rm tot}(0),\pex(0)) \right)
   \right]
 \ket =1.
 \label{ons-Jarz}
\end{equation}
This  is in the same form as the integral fluctuation theorem~\cite{seifert2012stochastic}, while its mathematical structure
is basically equivalent to the Jarzynski equality~\cite{PhysRevLett.78.2690}.

Because $S_*(E_{\rm tot}(0),\pex(0))$ is independent of $\hat V$,
we also have 
\begin{equation}
\bra 
 \exp
 \left[-  S_*(E_{\rm tot}({t_f}),\pex({t_f}))  \right]
 \ket =  \exp \left[-  S_*(E_{\rm tot}(0),\pex(0)) \right].
 \label{ons-Jarz-2}
\end{equation}
From $e^{-x} \ge 1-x$ for any real number $x$, we find
\begin{equation}
\bra S_*(E_{\rm tot}({t_f}),\pex({t_f}))  \ket \ge 
S_*(E_{\rm tot}(0),\pex(0)) ,
 \label{2nd}
\end{equation}
which is the second law of thermodynamics.

\subsection{Quasi-static processes}

When the time scale of change in $\pex(t)$ is much longer
than that of the relaxation time of thermodynamic quantities in
the system, such an operation is called {\it quasi-static operation}.
Explicitly, we assume that the relaxation time is one by appropriate
dimensionless forms, and we introduce a small parameter $\epsilon$
by which  we can express the protocol $\pex(t)$, $0 \le t \le {t_f}$,  as 
\begin{eqnarray}
\pex(t) =   \bar \pex(\epsilon t), \label{scale-V} \\
t_f  = \frac{\tau_f}{\epsilon},
\end{eqnarray}
where the functional form of $\bar \pex$ and $\tau_f$ are
independent of $\epsilon$.

Next, we take $\Delta$ such that $1 \ll \Delta \ll \epsilon^{-1}$ and define 
\begin{equation}
 \bar V_n \equiv \frac{1}{\Delta} \int_{n \Delta}^{(n+1)\Delta} dt V(t),
\label{vn-def}
\end{equation}
to express 
\begin{eqnarray}
  E_{\rm tot}({t_f})-E_{\rm tot}(0)
  &=&  \int_0^{{t_f}} dt V(t) \der{}{t}\bar \pex(\epsilon t),  \nonumber \\
  &=&  \sum_n \epsilon \Delta \bar V_n
  \bar \pex'(\tau_n)+O((\epsilon \Delta)^2), \nonumber \\
  &=&  \int_0^{\tau_f} d\tau \bar V(\tau)  \bar \pex'(\tau),
\end{eqnarray}
where $\tau_n=n\Delta \epsilon$ and $\bar V(\tau_n)\equiv \bar V_n$.
By applying the law of large numbers to (\ref{vn-def}),
we conclude that the  fluctuations of $\bar V(\tau)$ tend to zero in the limit
$\Delta \gg 1$, and, consequently,
$E_{\rm tot}({t_f})-E_{\rm tot}(0)$ takes a typical value with high probability in the quasi-static limit. Therefore, (\ref{ons-Jarz-2}) leads to the entropy conservation:
\begin{equation}
S_*(E_{\rm tot}(t_f),\pex(t_f)) = S_*(E_{\rm tot}(0),\pex(0)).
 \label{2nd-qs}
\end{equation}

\subsection{MSRJD formalism}

By introducing an auxiliary variable $\pi$, we can express (\ref{OM}) as
\begin{eqnarray}
\hspace{-1.5cm}  {\cal P}(\hat V|V(0),E_{\rm tot}(0), \hatpex  ) \nonumber \\
\hspace{-1.5cm}   = {\cal N}' \int \D \pi
 \exp\left\{ - \int_0^{{t_f}} dt L \left[\pi+ \frac{i}{2L} 
  \left( \der{V}{t}- L \der{S(E_{\rm tot}-\pex V, V)}{V} \right)
  \right]^2  \right\}  \nonumber \\
  \exp\left\{
    -\int_0^{{t_f}} dt \left[
      \frac{1}{4L}
      \left( \der{V}{t}- L \der{S(E_{\rm tot}-\pex V, V)}{V} \right)^2
                     \nonumber \right. \right.  \nonumber \\
 \left. \left.
      +
      \frac{1}{2}\dert{S(E_{\rm tot}-\pex V, V)}{V} 
      \right] \right\} ,
\label{OM-2}
\end{eqnarray}
which is further rewritten as
\begin{equation}
{\cal P}(\hat V|V(0),E_{\rm tot}(0), \hatpex  )
={\cal N}' \int \D\pi \exp\left[ I_{\rm MSR}  \right],
\end{equation}
where
\begin{eqnarray}
  \IMSR[V,\pi]=&& \int_0^{{t_f}} dt \biggl[
    -i \pi
    \left( \der{V}{t}- L \der{S(E_{\rm tot}-\pex V, V)}{V} \right)
   - L \pi^2  \nonumber \\
   && -  \frac{1}{2}\dert{S(E_{\rm tot}-\pex V, V)}{V} 
   \biggr] .
\label{eq:IE}
\end{eqnarray}

For any observable ${\cal O}$ that is a function of $\hat V$,
the expectation of ${\cal O}$ is expressed as 
\begin{eqnarray}
   \left\langle {\cal O} \right \rangle
  \equiv   \int \D \pi \D V \exp[\IMSR] {\cal O},
  \label{eq:expect}
\end{eqnarray}
where the normalization constant is absorbed into the integral
measure to satisfy $ \left\langle 1 \right \rangle =1 $.
The path integral representation (\ref{eq:expect}) is called
the MSRJD formalism.

In order to see a physical interpretation of the 
auxiliary variable $\pi$, we add a perturbation
$f_p(t)$  to the right-hand side of (\ref{ons}) and consider the
expectation of a quantity $A(t')$ at time $t'$ denoted by
$\bra A(t') \ket_{f_p(t)}$. For an infinitely small $f_p(t)$, we define
the response function $R_A(t',t)$ as
\begin{equation}
\bra A(t') \ket_{f_p(t)}-  \bra A(t') \ket= \int_0^{t'} dt R_A(t',t)f_p(t).
\end{equation}
By using (\ref{eq:expect}), we obtain
\begin{equation}
 R_A(t', t)= \bra  A(t') i\pi (t) \ket
\end{equation}
for $t' \ge t$. From this expression,
one may call $\pi$ the ``response variable''.
Furthermore, from (\ref{OM-2}), we also derive
\begin{equation}
  \bra  A(t') i\pi (t) \ket
 = \frac{1}{2L} \bra  A(t')
 \left( \der{V}{t}- L \der{S(E_{\rm tot}-\pex V, V)}{V} \right) \ket.
\label{minami}
\end{equation}

\section{Symmetry} 
\label{sec:symmetry}

To explore a symmetry property, we map the MSRJD formalism
to Hamilton mechanics by setting $p=  i\pi$ and $q=V$ in (\ref{eq:IE}).
We have 
\begin{equation}
\IMSR=- \int_0^{{t_f}} dt \left[{p} \der{q}{t}-H({p},{q})  \right] 
\label{eq:IE-z}
\end{equation}
with
\begin{equation}
H({p},{q})=L p^2 +L p \der{S}{q}-\frac{1}{2}\dert{S}{q} .
\end{equation}
For a more detailed discussion on mapping, see~\cite{minami2018spontaneous, justin1989quantum}. 
Next, we consider the Hamiltonian system with $H(q,p)$. 
From the Poisson bracket 
\begin{eqnarray}
\{q,p\}_{\rm PB}&=&1 , \\
\{q,q\}_{\rm PB}&=&\{p,p\}_{\rm PB}=0,
\end{eqnarray}
we obtain 
\begin{eqnarray}
\{S, p\}_{\rm PB}=& \der{S}{q}, \\
\{S,q\}_{\rm PB}=&0.
\end{eqnarray}
Therefore, $S$ is interpreted as the generator of the
following transformation:
\begin{eqnarray}
q  &\to q,  \label{eq:trans1}\\
p &\to p+ \eta \der{S}{q}, \label{eq:trans2}
\end{eqnarray}
where $\eta$ is a dimensionless infinitesimal transformation parameter.
Note here that the transformation is defined for the system
irrespective of symmetry. 

Here, we consider the variation of the MSRJD action (\ref{eq:IE})
for the transformation (\ref{eq:trans1}) and (\ref{eq:trans2}),
which is calculated as
\begin{eqnarray}
\delta \IMSR &=& -\eta \int^{t_f}_{0} dt \biggl[
 \der{q}{t}\der{S}{q}-L \left(\der{S}{q} \right)^2-2L  p \der{S}{q} \biggr] \\
&=& -\eta \int^{t_f}_{0} dt \biggl[
\der{S}{t}-L \left(\der{S}{q} \right)^2-2L p \der{S}{q}
\biggr], \label{eq:deltaIE}
\end{eqnarray}
where we have used
\begin{eqnarray}
\der{S}{t} =    \der{q}{t}\der{S}{q}.
\end{eqnarray}
We focus on restricted trajectories near equilibrium, i.e., the trajectories that satisfy 
\begin{eqnarray}
\der{S}{q}  = O(\epsilon), \\
p  = O(\epsilon).
\end{eqnarray}
For such restricted trajectories near the equilibrium, we find 
\begin{eqnarray}
\delta \IMSR 
&=-\eta \int^{t_f}_{0}dt \biggl[\frac{d S}{dt}+O(\epsilon^2) \biggr].
\end{eqnarray}
Because $\delta \IMSR $ is written as the time integration of the total
derivative of a function, we identify the symmetry for the
transformation given by (\ref{eq:trans1}) and (\ref{eq:trans2}).
Based on Noether's theorem that we review below,
we can claim that $S$ is the Noether invariant for the system
we study.

Now, we derive Noether's theorem for the action of the form
\begin{eqnarray}
\IMSR[p,q]=-\int^{t_f}_{0} dt \biggl[ p  \der{q}{t}-H(p,q)\biggr],
\end{eqnarray}
For a general transformation
\begin{eqnarray}
p &\to p+\eta \delta p, \\
q &\to q+\eta \delta q,
\end{eqnarray}
the variation of the action  is calculated as
\begin{eqnarray}
  \hspace{-0.5cm}  \delta \IMSR =-\eta \int^{t_f}_{0} dt \biggl[\frac{d}{dt}\biggl(p \delta q\biggr)+\delta p \biggl(\der{q}{t}-\frac{\partial H}{\partial p}\biggr)-\delta q \biggl(\der{p}{t}+\frac{\partial H}{\partial q}\biggr)\biggr].
\end{eqnarray}

Such an action is called symmetric when the variation $\delta I$ can be written
as the time integral of the total time derivative of some quantity $\Psi$: 
\begin{eqnarray}
\delta I = -\eta \int^{t_f}_{0} dt \frac{d \Psi}{dt}.
\end{eqnarray}
Then, we have the relation
\begin{eqnarray}
  \hspace{-0.7cm}   \int^{t_f}_{0} dt \biggl[\frac{d}{dt}\biggl(p \delta q\biggr)+\delta p  \biggl(\der{q}{t}-\frac{\partial H}{\partial p}\biggr)-\delta q \biggl(\der{p}{t}+\frac{\partial H}{\partial q}\biggr)\biggr]
=\int^{t_f}_{0} dt \frac{d \Psi}{d t}.
\end{eqnarray}
Because the second and the third terms vanish along solution trajectories
of the Hamilton equation, we obtain that
\begin{eqnarray}
\int^{t_f}_{0} dt \biggl[\frac{d}{dt}\biggl(\Psi - p\delta q \biggr)\biggr]=0
\end{eqnarray}
holds along the solution trajectories. 
Therefore, the quantity  $Q\equiv \Psi - p \delta q $ is the Noether invariant.

For the transformation given by (\ref{eq:trans1}) and (\ref{eq:trans2}),
the above quantities should be read as 
\begin{eqnarray}
\delta q &= 0, \\
\delta p &= \frac{d S}{d V}, \\
\Psi &=  S.
\end{eqnarray}
Therefore, the Noether invariant is the thermodynamic entropy:
\begin{eqnarray}
Q= S.
\end{eqnarray}

At the end of this section, we present another form of the action
$I_S$ in terms of the entropy $S$. Using the transformation from $V$ to $S$, we obtain
\begin{equation}
 I_S= \int_0^{{t_f}} dt \left[
    -i \theta 
    \left( \der{S}{t}- L \biggl(\der{S}{V}\biggr)^2 \right)
   - L \left(\der{S}{V} \right)^2\theta^2 -  \frac{1}{2}\dert{S}{V} 
   \right] ,
\label{eq:IS} 
\end{equation}  
where $\theta=(d V /d S) \pi $.
In this representation, the transformation is given as the uniform
translation of $\theta$:
\begin{eqnarray}
\theta &\to \theta +i \eta, \label{eq:transS1}\\
S &\to S.\label{eq:transS2}
\end{eqnarray}
Therefore, for trajectories satisfying $dS/dV=O(\epsilon)$, the action (\ref{eq:IS}) 
possesses the symmetry for this transformation.

\section{Identity} 
\label{sec:identity}

We set $\Ob=1$ in  (\ref{eq:expect}) and change the integral
variable $\pi$ as 
\begin{eqnarray}
\pi{'}(t) =\pi(t) +i\eta\der{S}{V}(t).
\end{eqnarray}
Then, we have 
\begin{eqnarray}
  1  &= \int  \D\pi{'}  \D V
    e^{\IMSR[V,\pi{'}]}  , \nonumber \\
&= \int  \D\pi \D V e^{\IMSR[V,\pi]+\delta \IMSR } , \\
&=1 + \langle \delta\IMSR \rangle .
\end{eqnarray}
We  thus obtain
\begin{eqnarray}
 \langle \delta\IMSR \rangle =0,
\label{eq:expect2}
\end{eqnarray}
where $\delta\IMSR$ is given by (\ref{eq:deltaIE}).

Substituting  (\ref{eq:deltaIE})
in (\ref{eq:expect2}),  we obtain
\begin{eqnarray}
\langle  S(t_f)-S(0) \rangle
=L \int^{t_f}_{0}dt
\biggl\langle\biggl( \der{S}{V} \biggr)^2\biggr\rangle
+2L \int^{t_f}_{0}dt
\left\langle  i \pi \der{S}{V} \right\rangle,
\label{eq:WTid}
\end{eqnarray}
which corresponds to the Ward-Takahashi identity in quantum field theory.
Because the right-hand side comes from the symmetry breaking terms,
(\ref{eq:WTid}) indicates that the increasing entropy arises
from the symmetry breaking. 
It should  also be noted that (\ref{eq:WTid}) can be derived by setting
$A= dS/dV$ in (\ref{minami}). 

Because in the quasi-static limit $\epsilon \to 0$,
$\langle S(t_f)-S(0) \rangle=O(\epsilon)$ holds from Noether's theorem,
we obtain
\begin{equation}
\biggl\langle\biggl( \der{S}{V} \biggr)^2\biggr\rangle
+2 \left\langle  i \pi \der{S}{V} \right\rangle=O(\epsilon^2).
\label{eq:ident-s-2}
\end{equation}

Next, we write the equation for $S$ as
\begin{eqnarray}
  \der{S}{t}= L \left( \der{S}{V} \right)^2 +
  L \dert{S}{V} + \der{S}{V} \cdot \sqrt{2L}\xi,
\end{eqnarray}         
where $\cdot$ is the ito product. Taking the expectation, we have
\begin{eqnarray}
  \biggl\langle \der{S}{t}  \biggr\rangle
  =L \bra \left( \der{S}{V} \right)^2 \ket +  L \bra \dert{S}{V} \ket.
\label{eq:ident-s}
\end{eqnarray}
By integrating this relation over the time interval $[0,t_f]$ 
and comparing the result with (\ref{eq:WTid}), we obtain
\begin{equation}
  \bra  i\pi (t) \frac{d S}{d V} \ket
 = \frac{1}{2} \bra  \dert{S}{V}   \ket.
\label{minami2}
\end{equation}
By combining (\ref{eq:ident-s-2}) with (\ref{minami2}), 
we have
\begin{equation}
\biggl\langle\biggl( \der{S}{V} \biggr)^2\biggr\rangle  
+ \bra  \dert{S}{V}   \ket =O(\epsilon^2),
\label{FRT}
\end{equation}
which corresponds to the fluctuation-response relation
at equal time.
To conclude, the symmetry breaking terms
lead to the fluctuation-response relation in the quasi-static
limit $\epsilon \to 0$.

For general non-quasi-static processes, from the second
law of thermodynamics  (\ref{2nd}), we find that
\begin{eqnarray}
\langle  S(t_f)-S(0) \rangle
=L \int^{t_f}_{0}dt
\biggl\langle\biggl( \der{S}{V} \biggr)^2\biggr\rangle
+L \int^{t_f}_{0}dt
\left\langle  \dert{S}{V} \right\rangle > 0.
\label{eq:WTid-f}
\end{eqnarray}
This means that the violation of the fluctuation-response
relation is related to the entropy production, which 
corresponds to the Harada-Sasa relation for Langevin
equations describing systems in contact with a heat
bath~\cite{harada2005equality,PhysRevE.73.026131}.
Furthermore, from the convexity of the entropy function,
we find that $d^2 S/dV^2 <0$. By substituting this relation, we obtain
\begin{eqnarray}
 \langle  S(t_f)-S(0) \rangle < 
L \int^{t_f}_{0}dt \biggl\langle\biggl( \der{S}{V} \biggr)^2\biggr\rangle,
\label{neq:TUR}
\end{eqnarray}
which represents an upper bound for the entropy production.
We also have a lower bound 
\begin{eqnarray}
L \int^{t_f}_{0}dt
\left\langle  \dert{S}{V} \right\rangle
< \langle  S(t_f)-S(0) \rangle. 
\label{neq:TUR2}
\end{eqnarray}
These inequalities may have some relevance to the trade-off relations,
including the thermodynamic uncertainty relation~\cite{barato2015thermodynamic, gingrich2016dissipation}.

\section{Concluding remarks}\label{remarks}

Although we have studied a rather simple example, our arguments are quite general. 
First, we write a Langevin equation for a complete set of slow
variables in a thermally isolated system following the Onsager theory.
Then, we can analyze the Langevin equation, similar to the specific
model we studied, and obtain the basically same result
for the symmetry property. 

However, one may be more familiar with Langevin equations describing stochastic processes in systems
with a heat bath. A typical and standard example of such a system
is a Brownian particle under an external control. For this case,
the entropy production becomes zero in  quasi-static processes. Thus,
an interesting future problem is to formulate such property as the result of
Noether's theorem. 

Next, we compare our result with those obtained in the previous
studies~\cite{sasa2016thermodynamic, sasa2019thermodynamical}.
For isolated quantum many body systems with an  external step protocol parameter,  the effective action possesses a symmetry for a uniform translation of the  canonical conjugate variable of the entropy~\cite{sasa2019thermodynamical}.   
The uniform transformation  
is rewritten as  (\ref{eq:transSY}) for a saddle point solution of the effective
action
\begin{eqnarray}
I_{QM}&=\int^{t_f}_{0} dt\biggl[-\frac{i}{\hbar}E-i \theta \dot{S} \biggr],\label{eq:ISSY}
\end{eqnarray}
where $E$, $S$ and $\theta$ are the energy, the entropy,
and the canonical conjugate variable to $S$, respectively.
Although the transformation is the same as that in our result,
our transformation cannot be rewritten as the non-uniform
time translation (\ref{eq:transSY}) that is found in~\cite{sasa2016thermodynamic}. 
The difference comes from the fact that the energy term
is absent in (\ref{eq:IS}). 

As a final remark, we notice that the action of the Langevin equation is a
different concept from the one of  the quantum dynamics. The action
of the Langevin equation comes from a transition probability,
while the action of the quantum dynamics comes from the time
evolution of a state vector. The action for the classical
mechanics is naturally understood from the quantum mechanics.
Therefore, we expect that there exists another type of action
which describes a quantum path integral in the thermodynamic
state space. Motivated by actions  (\ref{eq:IS})
and (\ref{eq:ISSY}), we conjecture that an action of such a
thermodynamic path integral formulation given in~\cite{sasa2019thermodynamical}
for our system can be written as
\begin{eqnarray}
  I[S,\dot{S}; \pex]
  =\frac{1}{2}I_S[S,\dot{S}; \pex]-\frac{i}{\hbar}\int^{t_f}_{0}dt
  E(S; \pex), \label{eq:Idissi}
\end{eqnarray}
which has the same symmetry as $I_S$ for the transformation
 (\ref{eq:transS1}) and (\ref{eq:transS2}). This leads to the conclusion that $S$ is the Noether invariant.
In addition, from the saddle point $\delta I / \delta S=0$,
we have
\begin{eqnarray}
\frac{d \theta}{dt}=\frac{2}{\hbar\beta}+O(\epsilon),\label{eq:EOM}
\end{eqnarray}
where we used $d S/ d V=O(\epsilon)$ in  quasi-static processes.
By using  (\ref{eq:EOM}), the transformation
 (\ref{eq:transS1}) could be rewritten as the non-uniform time translation
\begin{eqnarray}
t \to t + 2\eta \hbar \beta+O(\epsilon).
\end{eqnarray}
We leave the derivation of the action (\ref{eq:Idissi}) from the thermodynamic
path integral formulation of \cite{sasa2019thermodynamical} for the future work.

\ack
We thank Y.~Hidaka, M. Itami  and H.~Nakano for useful discussions.
The present study was supported by KAKENHI (Nos. 17H01148, 19H05496,
19H05795), by the Zhejiang Provincial Natural Science Foundation
Key Project (Grant No. LZ19A050001), and by NSF of China (Grant No. 11674283).

\end{document}